# Conceptual Level Design of Semi-structured Database System: Graph-semantic Based Approach

Anirban Sarkar
Department of Computer Applications
National Institute of Technology, Durgapur
West Bengal, India

*Abstract*—**This paper has proposed a Graph – semantic based conceptual model for semi-structured database system, called GOOSSDM, to conceptualize the different facets of such system in object oriented paradigm. The model defines a set of graph based formal constructs, variety of relationship types with participation constraints and rich set of graphical notations to specify the conceptual level design of semi-structured database system. The proposed design approach facilitates modeling of irregular, heterogeneous, hierarchical and non-hierarchical semi-structured data at the conceptual level. Moreover, the proposed GOOSSDM is capable to model XML document at conceptual level with the facility of document-centric design, ordering and disjunction characteristic. A rule based transformation mechanism of GOOSSDM schema into the equivalent XML Schema Definition (XSD) also has been proposed in this paper. The concepts of the proposed conceptual model have been implemented using Generic Modeling Environment (GME).**

***Keywords- Semi-structured Data; XML; XSD; Conceptual Modeling; Semi-structured Data Modeling; XML Modeling.***

## I. INTRODUCTION

The increasingly large amount of data processing on the web based applications has led a crucial role of semi-structured database system. In recent days, semi-structured data has become prevalent with the growing demand of such web based software systems. Semi-structured data though is organized in semantic entity but does not strictly conform to the formal structure to strict types. Rather it possess irregular and partial organization [1]. Further semi-structured data evolve rapidly and thus the schema for such data is large, dynamic, is not strict to type and also is not considered the participation of instances very strictly.

The eXtensible Markup Language (XML) is increasingly finding acceptance as a standard for storing and exchanging structured and semi-structured information over internet [12]. The Document Type Definition (DTD) or XML Schema Definition (XSD) language can be used to define the schema which describes the syntax and structure of XML documents [9]. However, the XML schemas provide the logical representation of the semi-structured data and it is hard to realize the semantic characteristics of such data. Thus it is important to devise a conceptual representation of semi-structured data for designing the information system based on such data more effectively. A conceptual model of semi-structured data deals with high level representation of the candidate application domain in order to capture the user ideas using rich set of semantic constructs and interrelationship thereof. Such conceptual model will separate the intention of designer from the implementation and also will provide a better insight about the effective design of semi-structured database system. The conceptual design of such system further can be implemented in XML based logical model.

To adopt the rapidly data evolving characteristics, the conceptual model of semi-structured data must support several properties like, representation of irregular and heterogeneous structure, hierarchical relations along with the non – hierarchical relationship types, cardinality, n – array relation, ordering, representation of mixed content etc. [13]. Beside these, it is also important to realize the participation constraints of the instances in association with some relation or semi-structured entity type. The participation of instances in semi-structured data model is not strict. In early years, Object Exchange Model has been proposed to model semi-structured data [2], where data are represented using directed labeled graph. The schema information is maintained in the labels of the graph and the data instances are represented using nodes. However, the separation of the structural semantic and content of the schema is not possible in this approach. In recent past, several researches have been made on conceptual modeling of semi-structured data as well as XML. Many of these approaches [3, 4, 5, 6, 7, 8] have been extended the concepts of Entity Relationship (ER) model to accommodate the facet of semi-structured data at conceptual level. The major drawbacks of these proposals are in representation of hierarchical structure of semi-structured data. Moreover, only two ER based proposals [7, 8] support the representation of mixed content in conceptual schema. In [7], a two level approach has been taken to represent the hierarchical relations. In first level the conceptual schema is based on extended concept of ER model and in second level, hierarchical organizations of parts of the overall conceptual schema are designed. In general, ER model are flat in nature [14] and thus unable to facilitate the reuse capability and representation of hierarchical relationship very efficiently. On the other hand, ORA-SS [11] proposed to realize the semi-structured data at conceptual level starting from its hierarchical structure. But the approach does not support directly the representation of no-hierarchical relationships and mixed content in conceptual level semi-structured data model.

Very few attempts have been made to model the semi-structured data using Object Oriented (OO) paradigm. ORA-SS [11] support the object oriented characteristic partially. The approaches proposed in [9, 10, 12] are based on UML. These approaches support object oriented paradigm comprehensively and bridge the gap between OO software design and semi-structured data schemata. However, the UML and extensions to UML represent software elements using a set of language elements with fixed implementation semantics (e.g. methods, classes). Henceforth, the proposed approaches using extension of UML, in general, are logically inclined towards implementation of semi-structured database system. This may not reflect the facet of such system with high level of abstraction to the user. In other word, semi-structured data model with UML extension cannot be considered as semantically rich conceptual level model. In [16] a graph semantic based web data model has been proposed and is appropriate for modeling structured web database system. The approach has not considered semi-structured characteristics of web databases.

In this paper, a graph semantic based conceptual model for semi-structured database system, called Graph Object Oriented Semi-Structured Data Model (GOOSSDM), has been proposed. The model is comprehensively based on object oriented paradigm. Among others, the proposed model supports the representation of hierarchical structure along with non-hierarchical relationships, mixed content, ordering, participation constraints etc. The proposed GOOSSDM reveals a set of concepts to the conceptual level design phase of semi-structured database system, which are understandable to the users, independent of implementation issues and provide a set of graphical constructs to facilitate the designers of such system. The schema in GOOSSDM is organized in layered approach to provide different level of abstraction to the users and designers. In this approach a rule based transformation mechanism also has been proposed to represent the equivalent XML Schema Definitions (XSD) from GOOSSDM schemata. The correctness of such transformation has been verified using the structural correlation mechanism described in [15]. Moreover, the concepts of proposed GOOSSDM have been implemented using Generic Modeling Environment (GME) [14] which is a meta-configurable modeling environment. The GME implementation can be used as prototype CASE tools for modeling semi-structured databases using GOOSSDM.

The preliminary version of this work has been published in [17] which has been now enriched and completed with comprehensive formalization and CASE tools.

## II. GOOSSDM: The Proposed Model

The GOOSSDM extends the object oriented paradigm to model semi-structured data. It contains all the details those are necessary to specify the irregular and heterogeneous structure, hierarchical and non-hierarchical relations, n – array relationships, cardinality and participation constraint of instances. The proposed data model allows the entire semi-structured database to be viewed as a Graph (V, E) in layered organization. At the lowest layer, each vertex represents an occurrence of an attribute or a data item, e.g. name, day, city etc. Each such basic attribute is to be represented as separate vertex. A set of vertices semantically related is grouped together to construct an *Elementary Semantic Group (ESG)*. So an ESG is a set of all possible instances for a particular attribute or data item. On next, several related ESGs are group together to form a *Contextual Semantic Group (CSG)*. Even the related ESGs with non-strict participations or loosely related ESGs are also constituent of related CSG – the constructs of related data items or attributes to represent one semi-structured entity or object. The edges within CSG are to represent the containment relation between different ESG in the said CSG. The most inner layer of CSG is the construct of highest level of abstraction or deeper level of the hierarchy in semi-structured schema formation. This layered structure may be further organized by combination of one or more CSGs as well as ESGs to represent next upper level layers and to achieve further lower level abstraction or higher level in the semi-structure data schema hierarchy. From the topmost layer the entire database appears to be a graph with CSGs as vertices and edges between CSGs as the association amongst them. The CSGs of topmost layer will act as roots of semi-structured data model schemata.

### A. Modeling Constructs in GOOSSDM

Since from the topmost layer, a set of vertices V is decided on the basis of level of data abstraction whereas the set of edges E is decided on basis of the association between different semantic groups. The basic components for the model are as follows,

A set of $t$ distinct *attributes* $A = \{a_1, a_2, ...., a_t\}$ where, each $a_i$ is an attribute or a data item semantically distinct.

(a) *Elementary Semantic Group (ESG)*: An elementary semantic group is an encapsulation of all possible instances or occurrences of an attribute, that can be expressed as graph *ESG (V, E)*, where the set of edges E is a null set $\varnothing$ and the set of vertices $V$ represent the set of all possible instances of an attribute $x_i \in A$. ESG is a construct to realize the elementary property, parameter, kind etc. of some related concern. Henceforth there will be set of $t$ ESGs and can be represented as $E_G = \{ESG_1, ESG_2, ...., ESG_t\}$. The graphical notation for the any ESG is *Circle*.

(b) *Contextual Semantic Group (CSG)*: A lowest layer contextual semantic group is an encapsulation of set of ESGs or references of one or more related ESGs to represent the context of one entity of semi-structured data. Let, the set of $n$ CSGs can be represented as $C_G = \{CSG_1, CSG_2, ..., CSG_n\}$. Then any lowest layer $CSG_i \in\subseteq C_G$ can be represented as a graph $(V_{Ci}, E_{Ci})$ where vertices $V_{Ci} \in E_G$ and the set of edges $E_{Ci}$ represents the association amongst the vertices. For any CSG, it is also possible to designate one or more encapsulated ESGs as *determinant vertex* which may determine an unordered or ordered set of instances of constituent ESGs or CSGs. The graphical notation for any CSG is *square* and determinant vertex is *Solid Circle*.

Composition of multiple CSGs can be realized in two ways. *Firstly*, the simple *Association* (Discussed in subsection II.B) may be drawn between two or more associated CSGs either of same layer or of adjacent layers to represent the non-hierarchical and hierarchical data structure respectively. The associated CSG will be connected using *Association Connector*. Those CSGs may share a common set of ESGs or referred ESGs.

*Secondly*, lower layer CSGs may maintain an *Inheritance* or *Containment* relationship (Discussed in subsection 2.B) with the adjacent upper layer CSG to represent the different level of abstraction. Thus, the upper layer CSG can be formed by inheritance or composition of one or more lower layer CSGs along with encapsulation of zero or more related ESGs or reference of ESGs. Then any upper layer $CSG_i \in C_G$ can be represented as a graph $(V_{Ci}, E_{Ci})$ where vertices $V_{Ci} \in C_G \cup E_G \cup Reference\ (E_G)$ and the set of edges $E_{Ci}$ represents the association amongst the vertices.

*(c) Annotation*: Annotation is a specialized form of CSG and can be expressed as $G(V, E)$, where $|V| = 1$ and $E = \varnothing$. Annotation can contain only text content as tagged value. Annotation can be containment in or associated with any other CSG. Further the cardinality constraint for Annotation construct is always *1:1* and ordering option can be *1* or *0*, where *1* means content will be in orderly form with other constituent ESG and *0* means text content can be mixed with other constituent ESGs. The annotation construct will realize the document-centric semi-structured data possibly with mixed content. This concept is extremely important for mapping semi-structured data model in XML. Graphically Annotation can be expressed using *Square with Folded Corner*.

The summary of GOOSSDM constructs and their graphical notations have been given in Table I.

TABLE I. SUMMARY OF GOOSSDM CONSTRUCTS AND THEIR GRAPHICAL NOTATIONS

| GOOSSDM Constructs | Description | Graphical Notation |
|---|---|---|
| ESG | Elementary Semantic Group | 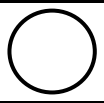 |
| Determinant ESG | Determinant vertex of any CSG which will determine the other member vertices in the CSG | 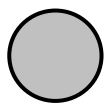 |
| CSG | Contextual Semantic Group | 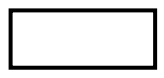 |
| Annotation | Specialized form of CSG. Contain only Text Content. | 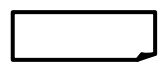 |
| Association Connector | Connect multiple associated CSGs | 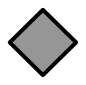 |

## B. Relationship Types in GOOSSDM

The proposed GOOSSDM provides a graph structure to represent semi-structured data. The edges of the graph represent relationships between or within the constructs of the model. In the proposed model, *four* types of edges have been used to represent different relationships. The type of edges and their corresponding meanings are as follows,

*(a) Containment:* Containments are defined between encapsulated ESGs including determinant ESG and parent CSG, or between two constituent CSGs and parent CSG, or between CSG and referential constructs. The Containment relationship is constrained by the parameters tuple $<p, \theta>$, where $p$ determines the participation of instances in containment and $\theta$ determines the ordering option of constituent ESGs or CSGs. With any CSG, this represents a bijective mapping between determinant ESG and other ESGs or composed CSG with participation constraint. Graphically association can be expressed using *Solid Directed Edge* from the constituent constructs to its parent labeled by constraint specifications. The possible values for $p$ are as follows,
*(i) 1:1* – Represents ESG with mandatory one instantiation or total participation in the relationship and is default value of $p$.
*(ii) 0:1* – Represents ESG with optional one instantiation in the relationship.
*(iii) 1:M* – Represents ESG with mandatory multiple instantiation in the relationship.
*(iv) 0:M* – Represents ESG with optional multiple instantiation in the relationship.
*(v) 0:X* – Represents ESGs with optional exclusive instantiation in the relationship. If a CSG contain single such ESG then it will act like *0:1* option. Otherwise one such ESG will optionally instantiate among all ESGs with $p$ value 0:X.
*(vi) 1:X* – Represents ESG with mandatory exclusive instantiation in the relationship. If a CSG contain single such ESG then it will act like 1:1 value option. Otherwise it is mandatory that one such ESG will instantiate among all ESGs with $p$ value 1:X.

The possible values for $\theta$ are as follows,
*(i) 1* – Represents that for any CSG, the constituent ESGs and CSGs are ordered and order must be maintained from left to right in the list of ESGs with $\theta$ value 1.
*(ii) 0* – This is default value of $\theta$ and represents that for any CSG, the constituent ESGs and CSGs are not ordered.

*(b) Association:* Associations are defined between related CSGs of same layer of adjacent layers. The Association relationship is constrained by the parameters tuple $<P, \theta>$, where $P$ determines the cardinality of Association and $\theta$ determines the ordering option of associated CSGs. Graphically association can be expressed using *Solid Undirected Edge*. Any CSG wish to participate in association will be connected with association relationship. On next, multiple associated CSGs will be connected through *Association Connector*. For semi-structured it is sometime important to have specific context of some association. Such context can be represented using *Associated CSG* defined on *Association Connector*. Association Connector facilitates the n – array relationship. Graphically association can be expressed using *Solid Undirected Edge*, Association connector can be expressed using *Solid Diamond* and Associated CSG can be connected with Association Connector using *Dotted Undirected Edges* with Participation constraint specifications. The values for $P$ can be *1:1* or *0:1* or *1:N* or *0:N* or *0:X* or *1:X* with corresponding meaning and possible values for $\theta$ can be *1* or *0* with corresponding meaning.

*(c) Link:* Links are used to represent the inheritance relationships between two CSGs. Graphically link can be expressed using *Solid Directed Edge with Bold Head* from the generalized CSG to the specialized one.

*(d) Reference:* In semi-structured data model, it is important to represent the symmetric relationship between ESGs or CSGs. Reference can be used to model such concepts. Reference relations are defined either between ESG and referred ESG or between CSG and referred CSG. Graphically reference can be expressed using *Dotted Directed Edge.*

The summary of GOOSSDM relationship types and their graphical notations have been given in Table II.

TABLE II. SUMMARY OF GOOSSDM RELATIONSHIP TYPES AND THEIR GRAPHICAL NOTATIONS

| GOOSSDM Relationships | Description | Graphical Notation |
|---|---|---|
| Containment | Defined between Parent CSG and constituent ESGs and CSGs | <p, θ> (solid line with open arrowhead) |
| Association | Defined between CSGs of same layer or adjacent layers. | <P, θ> (solid line) |
| CSG Association | Defined between association and associated CSG | <P, θ> (dash-dot line) |
| Link | Defined between two adjacent layer parent CSG and inherited CSG | (solid line with filled arrowhead) |
| Reference | Defined either between ESG and referred ESG or between CSG and referred CSG. | (dotted line with open arrowhead) |

## III. TRANSFORMATION OF GOOSSD INTO XSD

In general, the proposed GOOSSDM can be useful to realize the semi-structured data schema at conceptual level. The logical structure of such schema can be represented using the artifacts of XSD. Moreover, XSD is currently the de facto standard for describing XML documents. An XSD schema itself can be considered as an XML document. *Elements* are the main building block of any XML document. They contain the data and determine the elementary structures within the document. Otherwise, XSD also may contain sub-element, attributes, complex types, and simple types. XSD schema elements exhibit hierarchical structure with single root element.

A systematic rule based transformation of GOOSD schemata to XSD is essential to express the semi-structured data at logical level more effectively. For the purpose, a set of rules have been proposed to generate the equivalent XSD from the semantic constructs and relationship types of a given GOOSSDM schemata. Based on the concepts of GOOSSDM constructs and relationship types the transformation rules are as follows,

*Rule1:* An ESG will be expressed as an *element* in XSD. For example, $ESG_{City}$ can be defined on attribute *City* to realize Customer city. Any DESG construct must be expressed with typed *ID* in XSD. The equivalent representation in XSD can be as given in Figure 1.

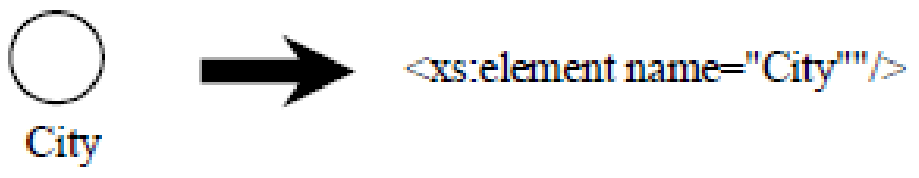

Figure1. Representation of ESG

*Rule 2:* A CSG will be expressed as a *complexType* in XSD. For example, $CSG_{Customer}$ can be defined to realize the detail of Customer. The equivalent representation in XSD can be as given in Figure 2.

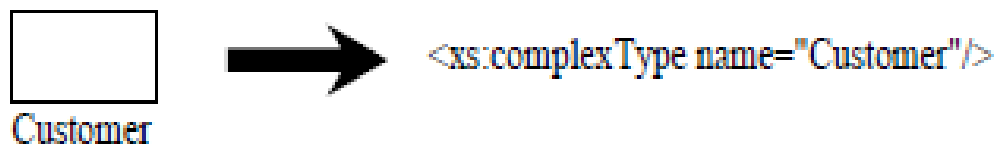

Figure 2. Representation of CSG

*Rule 3:* Any Annotation construct will be expressed as a *complexType* in XSD with suitable *mixed* value. On containment to other CSG, if $\theta$ value is 0 then it will be treated as mixed content in the resulted XML document. Otherwise if $\theta$ value is 1 then it will be treated as annotation text in resultant XML document in orderly form. An example of annotation constructs has been shown in Figure 3.

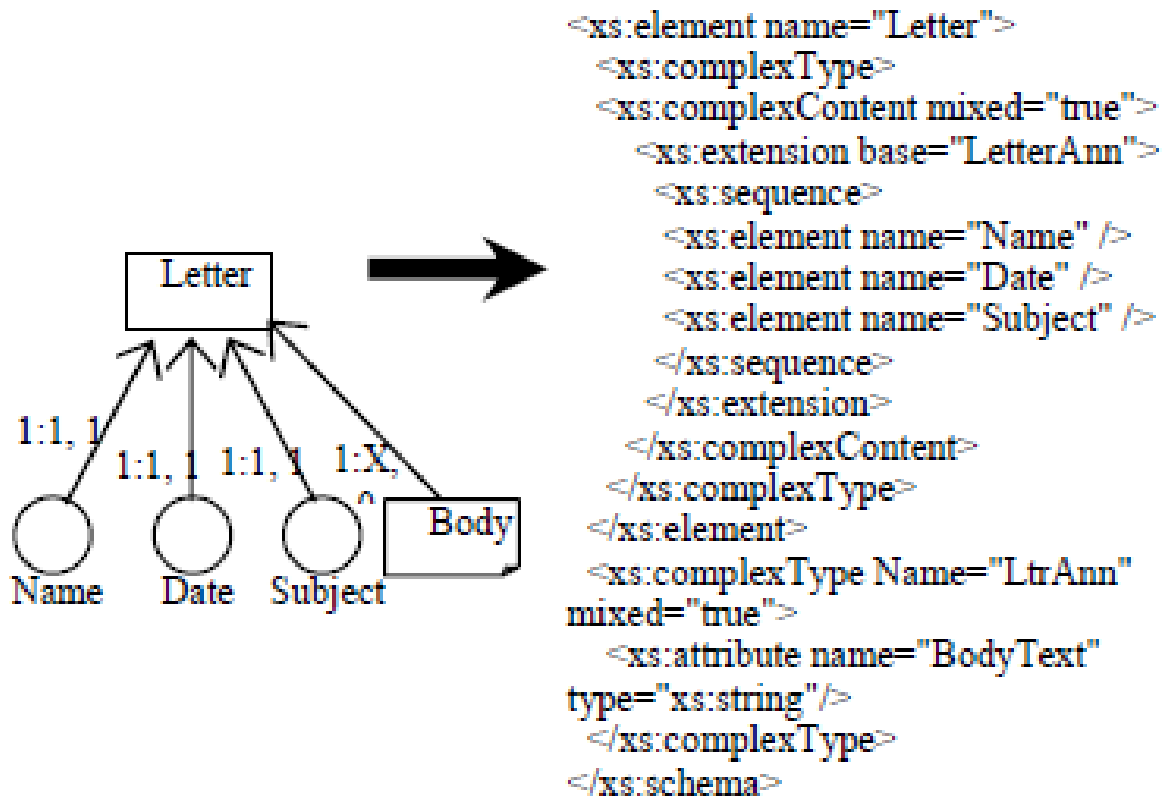

Figure 3. Representation of Annotation Construct

*Rule 4:* A Reference of ESG and CSG will be expressed as a *complexType* in XSD. For example, a reference of $ESG_{City}$ can be defined on attribute *City* to realize a referential attribute on Customer city. The equivalent representation in XSD can be as given in Figure 4.

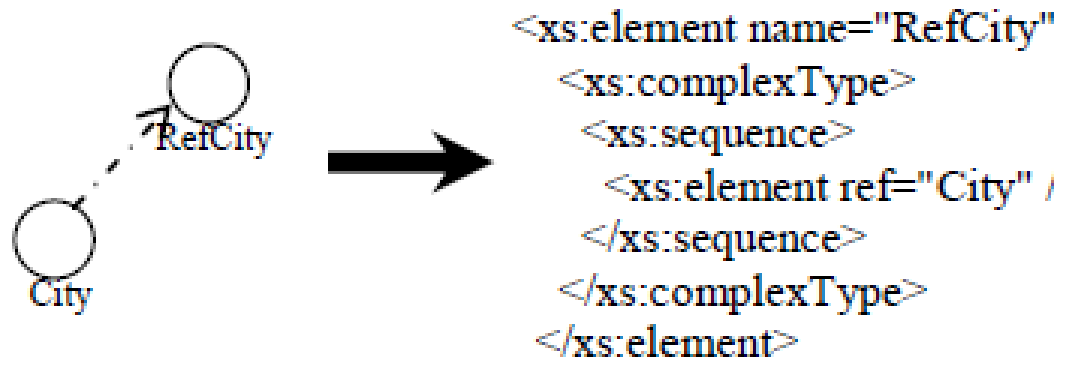

Figure 4. Representation of Reference of ESG.

*Rule 5:* CSGs of topmost layer will be treated as root in XSD declaration.

*Rule 6:* Any lowest layer CSG with containment of some ESGs will be expressed as a *complexType* with *elements* declaration in XSD. Further the participation constraint (*p value* in GOOSSDM concept) can be expressed using *minOccurs* and *maxOccurs* attribute in XSD. The ordering constraint ($\theta$ *value* in GOOSSDM concept) can be expressed using compositor type of XSD. If $\theta$ *value* is 1 then compositor type will be *sequence* otherwise *all*. For ordered set ESGs, the order will be maintained from *left to right*. If any subset of ESGs contains the *p value X:1* then those ESGs will be composite using *choice* compositor type in XSD. An example of XSD representation of lowest layer CSG has been shown in Figure 5.

*Rule 7:* Any upper layer CSG with containment of ESGs, reference of ESGs and adjacent lower layer CSGs will be expressed as a *complexType* in XSD. An example of XSD representation of upper layer CSG with containment relation has been shown in Figure 6.

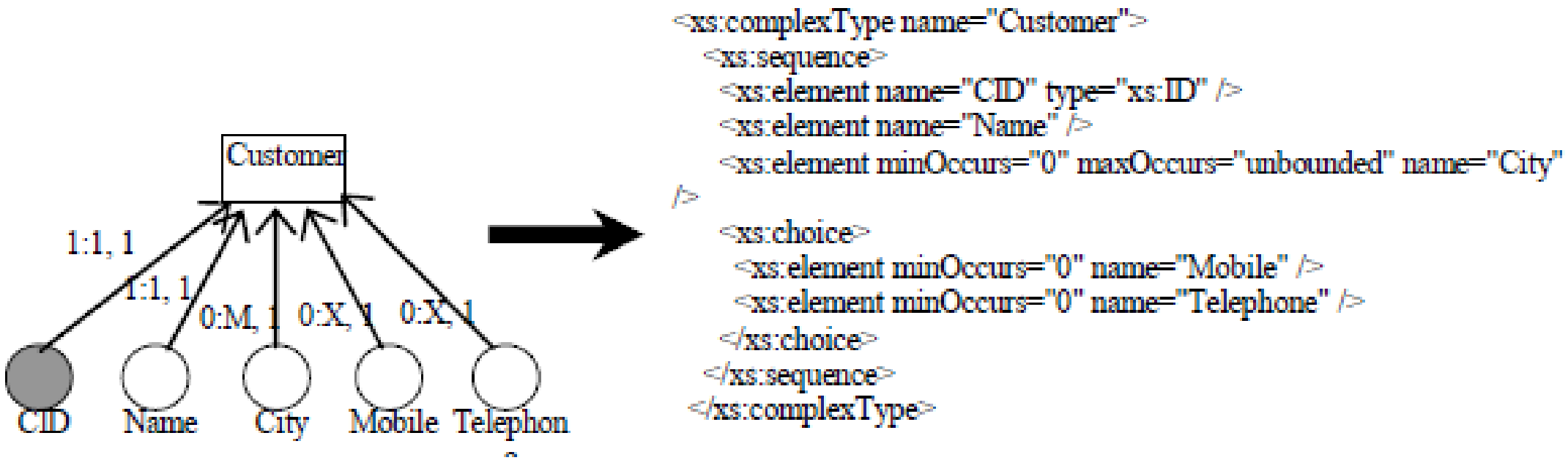

Figure 5. Representation of Lower layer CSG

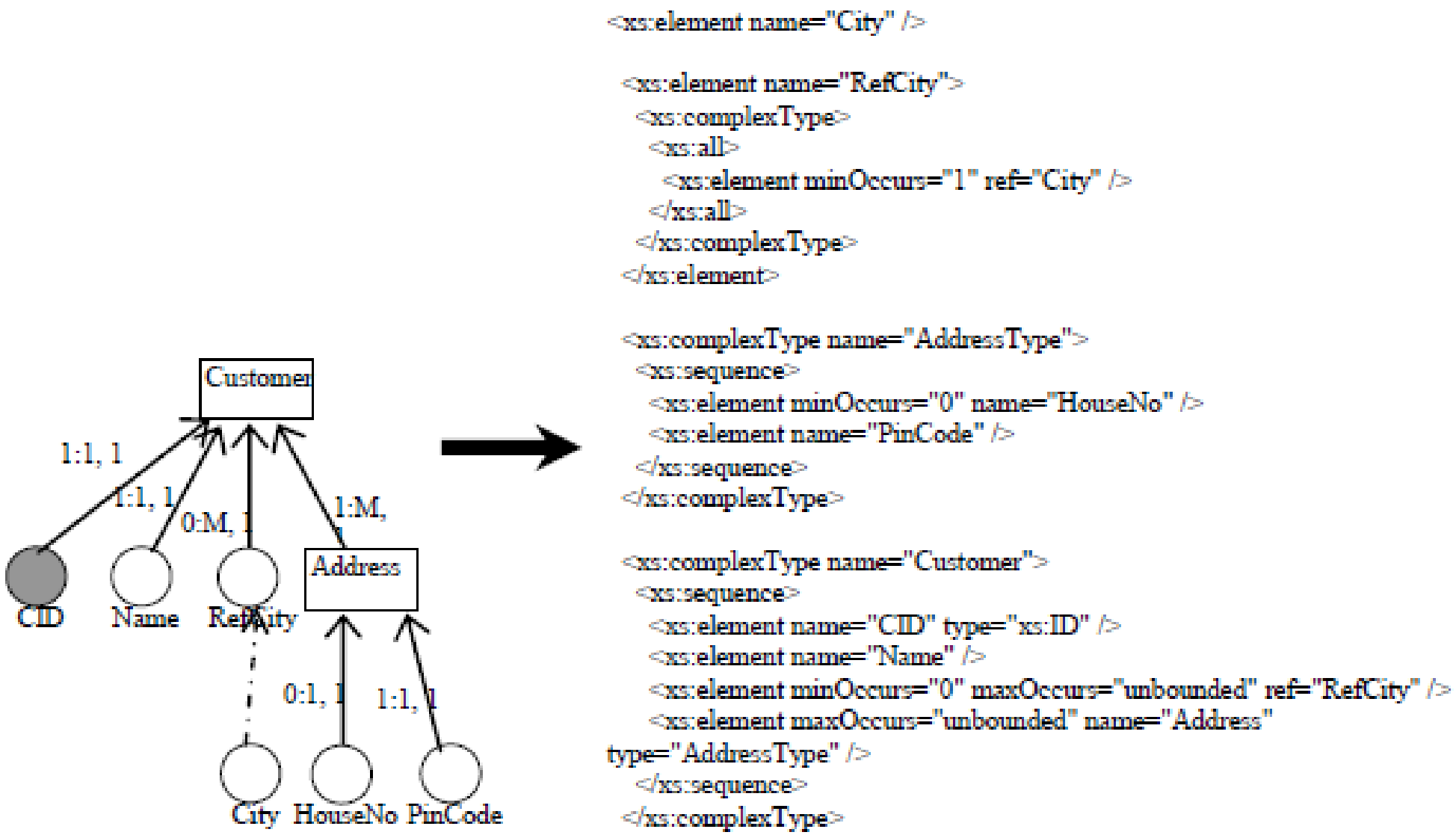

Figure 6. Representation of Upper layer CSG.

*Rule 8:* Any upper layer CSG with *Link* relationship with adjacent lower layer CSGs will be expressed as a *complexType* with inheritance in XSD. Upper layer CSG will be the child of lower layer CSG. An example of XSD representation of upper layer CSG with inheritance with adjacent lower layer CSG has been shown in Figure 7.

*Rule 9:* Any upper layer CSG with *Association* relationship with adjacent lower layer CSGs will be expressed as a *complexType* with nesting in XSD. Upper layer CSG will be treated as root element.

*Rule 10: Association* relationship between any two CSGs in the same layer will be expressed as a *complexType* with nesting in XSD. Rightmost CSG will be treated as the root element and on next nesting should be done in *right to left* order of the CSG in the same layer.

*Rule 11:* N – array *Association* relationship within a set of CSGs spread over several layer will be expressed as a *complexType* with nesting in XSD. Topmost layer CSG will be treated as the root element in XSD. Then, in the adjacent lower layer the rightmost CSG should be treated as nested element within the root element. Further the nesting should be done in right to left order of the CSG in the same layer and on next moving on the adjacent lower layers.

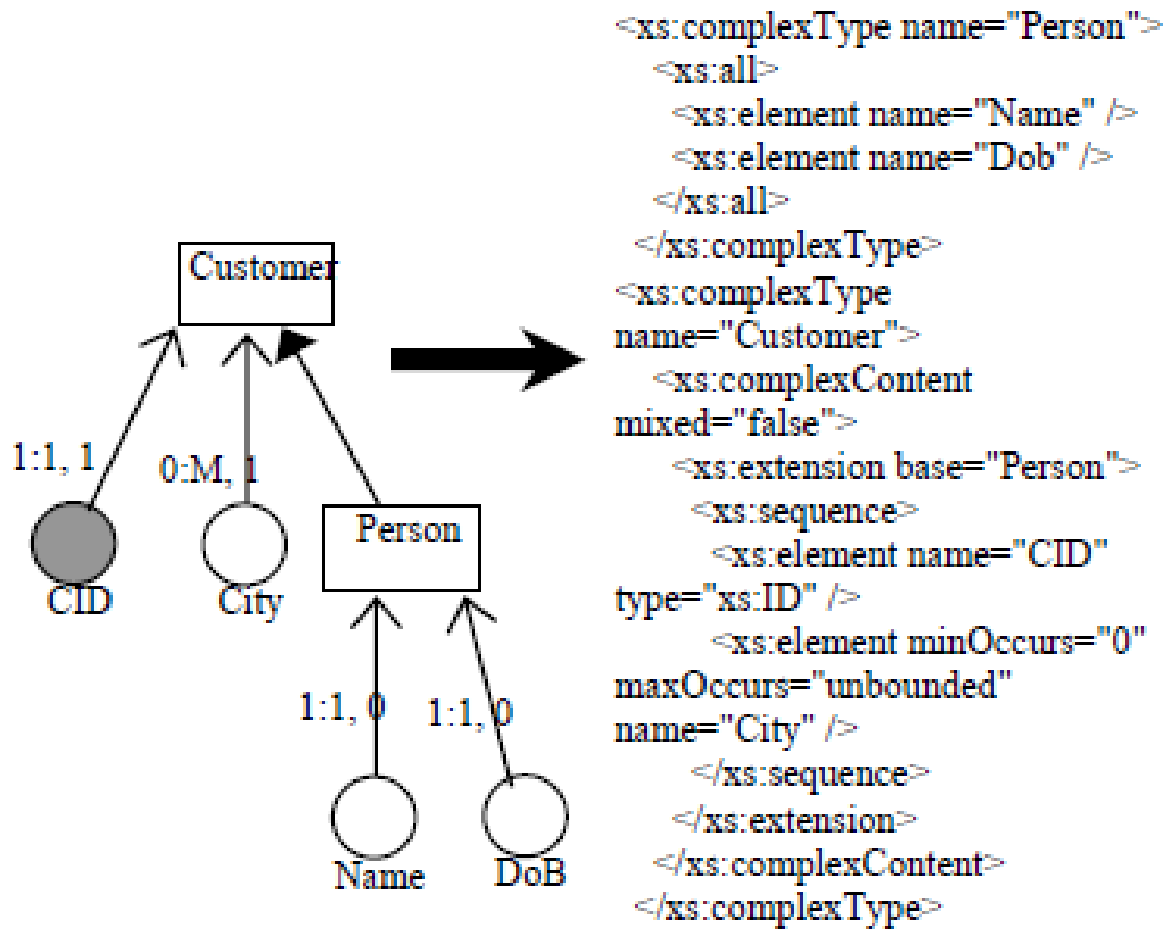

Figure 7. Representation of Link Relationship

*Rule 12:* With several Association relationships (composition of n – array and simple relationship) within a set of CSGs spread over several layer will be expressed as a complexType with nesting in XSD. Topmost layer CSG will be treated as the root element in XSD. Then, if available, the directly associated CSGs in each adjacent lower layer will be nested till it reaches

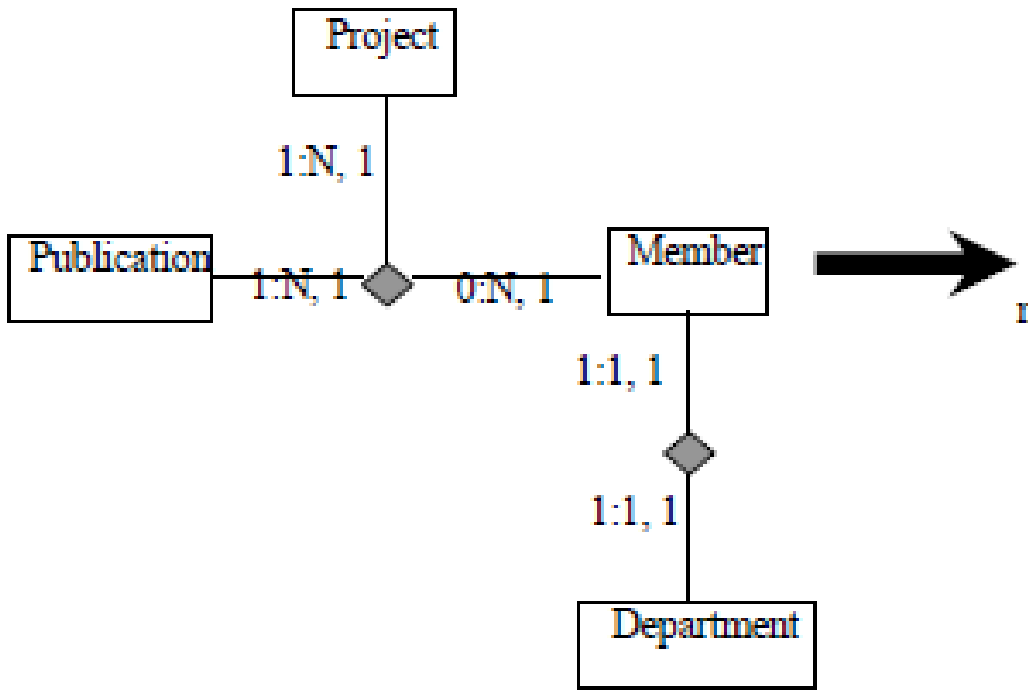

```
<xs:complexType name="Project">
  <xs:sequence maxOccurs="unbounded">
    <xs:element name="PID" type="xs:ID"/>
    <xs:element name="PTitle"/>
    <xs:element minOccurs="0" maxOccurs="unbounded" name="Member">
      <xs:complexType>
        <xs:sequence>
          <xs:element name="MID" type="xs:ID"/>
          <xs:element name="MNAME"/>
          <xs:element name="Department">
            <xs:complexType>
              <xs:sequence>
                <xs:element name="DID"/>
                <xs:element name="DName"/>
              </xs:sequence>
            </xs:complexType>
          </xs:element>
          <xs:element minOccurs="0" maxOccurs="unbounded"
name="Publications">
            <xs:complexType>
              <xs:all>
                <xs:element name="Num"/>
                <xs:element name="Title"/>
              </xs:all>
            </xs:complexType>
          </xs:element>
        </xs:sequence>
      </xs:complexType>
    </xs:element>
  </xs:sequence>
</xs:complexType>
```

Figure 8. Representation of Associated CSGs spread over several layers (ESG layer is hidden).

to the lowermost layer of available associated CSG. On next, the CSGs of adjacent lower layer of the root element will be nested from right to left order in the same layer along with the nesting of directly associated CSGs (if available) in each corresponding adjacent layers. An example has been shown in Figure 9 for XSD representation of GOOSSDM schemata where associated CSGs are spread over three layers and contain both n – array and simple associations.

```
<Patient>
 <name>Patient 1</name>
 <Visit>
        <Date>10-JAN-2009</Date>
        <Doctor>
                <RegID>1234</RegID>
                <DName>Dr. P. Roy</DName>
        </Doctor>
        <Dept><DID>1</DID><DeptName>General</DeptName>
            <Hospital><Name>Hospital A</Name> </Hospital>
        </Dept>

        <Date>15-MAR-2010</Date>
        <Doctor>
                <RegID>4321</RegID>
                <DName>Dr. T. De</DName>
        </Doctor>
        <Clinic><Name>Clinic B</Name></Clinic>
 </Visit>

 <name>Patient 2</name>
 <Visit>
        <Date>12-SEPT-2009</Date>
        <Doctor>
                <RegID>4321</RegID>
                <DName>Dr. T. De</DName>
        </Doctor>
        <Clinic><Name>Clinic D</Name></Clinic>
 </Visit>
</Patient>
```

Figure 9. Irregular Structure in Visit Records XML

## IV. CASE STUDY

Let consider an example of *Visit Record* of *Patient* where a *Patient* can visit to a *Doctor* either at *Hospital Department* or at *Clinic* [7]. Any patient can visit several times to different doctors. The Figure 9 shows an irregularly structured XML representation of visit records of two patients. *Patient 1* visited twice to two different *Doctors*, one at *Hospital Department* and another at *Clinic*. *Patient 2* visited once to one *Doctor* common to *Patient 1* but at different Clinic. All though in the document, the *Date* of *Visit*, *Doctor* and option of *Hospital Department* and *Clinic* are in order. The XML document of Figure 9 represents the semi-structured data for such *Visit Record* database. The suitable GOOSSDM schemata for such data and its equivalent XSD have been shown in Figure 10. The equivalent XSD of GOOSSDM schemata of Figure 10 can be generated using the rules described in Section III.

## V. CORRECTNESS OF GOOSSDM TRANSFORMATION

The set of proposed transformation rules described in Section III facilitates the systematic transformation of conceptual level semi-structured data model like GOOSSDM to the equivalent XSD in logical level. The correctness of the model transformation can be proved using the structural correspondence approach described in Narayanan *et al* [15]. In every model transformation, there is a correlation or correspondence between parts of the input model and parts of the output model. One can specify these correlations in terms of the abstract semantics of the source and target model constructs. The approach of Narayanan et al. describes that, if a transformation has resulted in the desired output models, there will be a verifiable structural correspondence between the source and target model instances that is decidable. Moreover, the transformation can be accepted as correct, if a node in the source model and its corresponding node in the target model satisfy some correspondence conditions.

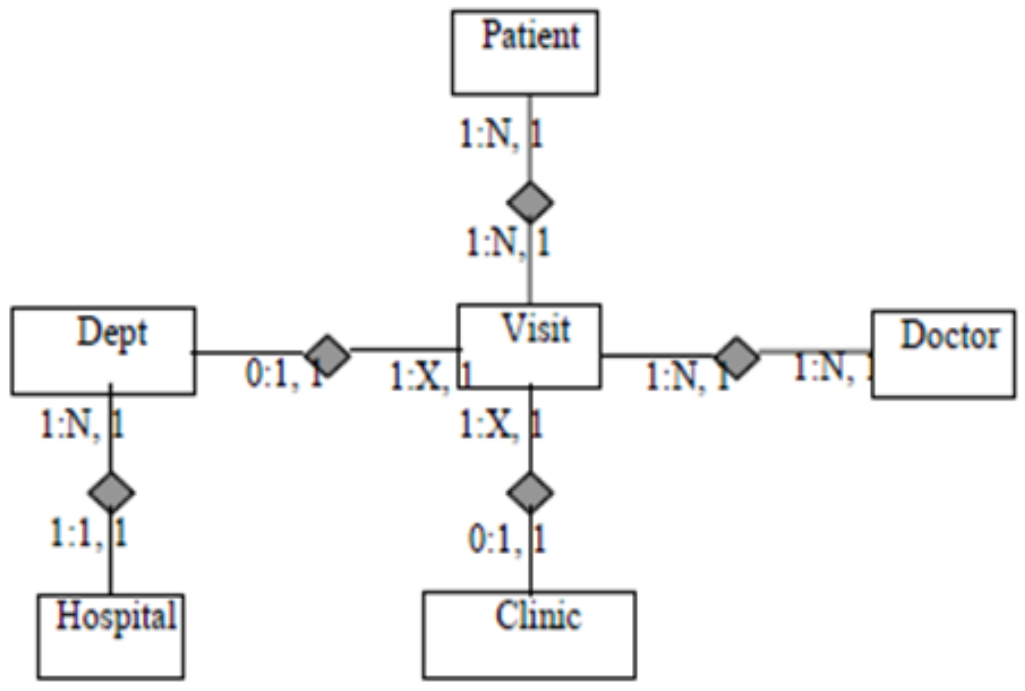

```
<xs:element name="Patient">
 <xs:complexType>
  <xs:sequence maxOccurs="unbounded">
   <xs:element name="name" />
   <xs:element maxOccurs="unbounded" name="Visit">
    <xs:complexType>
     <xs:sequence maxOccurs="unbounded">
      <xs:element name="Date" />
      <xs:element name="Doctor">
       <xs:complexType>
        <xs:sequence>
         <xs:element name="RegID" />
         <xs:element name="DName" />
        </xs:sequence>
       </xs:complexType>
      </xs:element>
      <xs:choice>
       <xs:element minOccurs="0" name="Dept">
        <xs:complexType>
         <xs:sequence>
          <xs:element name="DID" type="xs:ID" />
          <xs:element name="DeptName" />
          <xs:element name="Hospital">
           <xs:complexType>
            <xs:sequence>
             <xs:element name="Name" />
            </xs:sequence>
           </xs:complexType>
          </xs:element>
         </xs:sequence>
        </xs:complexType>
       </xs:element>
       <xs:element minOccurs="0" name="Clinic">
        <xs:complexType>
         <xs:sequence>
          <xs:element name="Name" />
         </xs:sequence>
        </xs:complexType>
       </xs:element>
      </xs:choice>
     </xs:sequence>
    </xs:complexType>
   </xs:element>
  </xs:sequence>
 </xs:complexType>
</xs:element>
```

Figure 10. Corresponding GOOSSDM schemata and Equivalent XSD of Figure 9

In case of GOOSSDM, the meta-model level identifiable correspondence structures are listed in Table III and the table can be treated as the *look-up table* for the cross links between the *source model* (GOOSSDM) and *target model* (XSD).

Further, the proposed set of rules will realize the correctness condition in model transformation. In our proposed approach, the correspondence rules must ensure that semantics and syntax for every constructs in the GOOSSDM model and its instance being transformed into the XSD model elements.

Several examples have been illustrated for the proposed transformation rules to verify the correctness of correspondence mapping of GOOSSDM schemata to the equivalent XSD.

TABLE III. LOOK-UP TABLE FOR STRUCTURAL CORRESPONDANCE

| **GOOSSDM Constructs** | **Graphical Notation** | **Equivalent XSD Representation** |
|---|---|---|
| ESG | | *xs:Element* |
| Determinant ESG | | *xs:ID* |
| CSG | | *xs:complexType* |
| Annotation | | *xs:complexType* with suitable *Mixed* value |
| Association Connector | | *xs:complexType* with nesting from right to left order |
| Containment | <p, θ> | *xs:complexType* with *element* declaration |
| Association | <P, θ> | *xs:complexType* with nesting |
| CSG Association | <P, θ> | *xs:complexType* with nesting from right to left order |
| Link | | *xs:extension* declaration |
| Reference | | *ref* declaration in *xs:element* |
| P value | 1:1 or 0:1 or 1:N or 0:N or 0:X or 1:X | *minOccurs* and *maxOccurs delcarations* |
| θ value | 1 or 0 | *Compositor* type : *all* or *sequence* |

## VI. FEATURES OF GOOSSDM

The proposed GOOSSDM is an extension of comprehensive object oriented model for Semi-structured Database System and which can be viewed as a Graph (V, E) in layered organization. It contains set of semantically enriched constructs and relationship types to describe all the details those are necessary to specify the artifacts of the system containing semi-structured data. Moreover, using proposed set of rules, the proposed model schemata can be systematically transformed into equivalent XSD, which represents the logical schema for semi-structured data. Apart from these, one of the major advantages of the model is that it defines each level of structural detail on the constructs which are independent of the implementation issues. Moreover, the graph structure maintains the referential integrity inherently. The features of the proposed model are as follows,

*(a) Explicit Separation of structure and Content:* The model provides a unique design framework to specify the design for the semi-structured database system using semantic definitions of different levels (from elementary to composite) of data structure through graph. The model reveals a set of structures like ESG, CSG, Annotation, Association Connector etc. along with a set of relationships like Containment, Association, Link, Reference etc. between the structures, which are not instance based or value based. So, the nature of contents that corresponded with the instances and the functional constraint on the instances has been separated from the system's structural descriptions.

*(b) Abstraction:* In the proposed GOOSSDM, the concepts of layers deploy the abstraction in semi-structured data schema. The upper layer views will hide the detail structural complexity from the users. Such a representation is highly flexible for the user to understand the basic structure of semi-structure database system and to formulate the alternative design options.

*(c) Reuse Potential:* The proposed model is based on Object oriented paradigm. It is supported with inheritance mechanism using the *Link* relationship. Henceforth, there is no binding in the model to reuse some CSG constructs of any layer. On reuse of CSG, the specialized CSG must be shown in adjacent upper layer of the parent CSG. Moreover, lowest layer ESG or lower layer can be shared and reused with different CSGs of the upper layers using *Containment* relationship.

*(d) Disjunction Characteristic:* The instances of semi-structured data schema are likely to be less homogeneous than structured data. Disjunction relationships facilitate the possibility of non-homogeneous instances. The proposed GOOSSDM supports disjunction relationship using the participation constraint attribute *p* or *P* (by setting *p or P* value either *0:X* or *1:X*). The Containment relationships between constituent ESGs or CSGs with the parent CSG can be disjunctive or Association relationships between two or more CSGs can be disjunctive. Figure 5 and Figure 10 respective explain such disjunctions.

*(e) Hierarchical and Non-hierarchical Structure:* The proposed model explicitly supports both hierarchical and non-hierarchical representation in semi-structure data modeling at conceptual level. Associated CSGs of different or same layers form the hierarchical or non-hierarchical structure in semi-structured data model. At the logical level modeling of semi-structured data using XSD supports only hierarchical structure. For the purpose, the set of rules have been proposed to transform more generous conceptual level schema to hierarchical logical schema.

*(f) Ordering:* Ordering is one important concept in modeling of semi-structured data. One or more attributes or relationships in semi-structured data schema can be ordered. Our proposed model supports ordering in two ways using the relationship ordering constraint attribute $\theta$. Firstly, the ordering may be enforced between parent CSG and any set of constituent ESGs and CSGs by specifying the $\theta$ value on containment relationship. Secondly, the ordering can be enforced on the any set of Association relationships within CSG.

*(g) Irregular and Heterogeneous structure:* By characteristic the semi-structured data is irregular and heterogeneous. The proposed GOOSSDM supports disjunction characteristic, ordering and representation of both hierarchical and non-hierarchical structure in the same schema. With all these facets, the proposed model can efficiently model the irregular and heterogeneous semi-structured data. Modeling of irregular structure using GOOSSDM has been shown in Figure 10.

*(h) Participation constraint:* Instances participations in the semi-structured data schema are not followed strictly. Participations of instances can be optional or mandatory or even exclusive for such schema. This can affect the participation of constituent ESGs and CSGs in the parent CSG or may affect the participation of CSGs in some association relationship either of simply type or n-array type. All these participation constraint can be modeled in proposed GOOSSDM by specifying the value for participation constraint attribute *p* or *P*.

*(i) Document-centric and Mixed Content:* In real world, document texts are mixed with semi-structured data. The feature is more important and frequent in XML documents. Thus it is an essence that, the conceptual model for semi-structured data must support modeling of such feature. In the proposed model, the *Annotation* construct facilitates to model document centric design of semi-structured data at conceptual level. Moreover, the modeling of the *Annotation* construct in the GOOSDM schema allows the instances of CSG and ESG to be mixed with the text content. The presence of this construct along with the other defined constructs and relationships, the proposed GOOSSDM is also capable to model XML document at conceptual level.

## VII. IMPLEMENTATION OF GOOSSDM USING GME

The Generic Modeling Environment (GME) provides meta-modeling capabilities and where a domain model can be configured and adapted from meta-level specifications (representing the Conceptual modeling) that describe the domain concept. It is common for a model in the GME to contain several numbers of different modeling elements with hierarchies that can be in many levels deep. The GME supports the concept of a viewpoint as a first-class modeling construct, which describes a partitioning that selects a subset of conceptual modeling components as being visible.

Moreover, GME support the programmatic access of the metadata of GME models. Most usual techniques for such programmatic access is to write GME interpreter for some metamodel. The interpreter will be able to interpret any domain model based on that predefined metamodel. GME interpreters are not standalone programs, they are components (usually Dynamic Link Libraries) that are loaded and executed by GME upon a user's request. Most GME components are built for the Builder Object Network (BON), an inbuilt framework in GME and provide a network of C++ objects. Each of these represents an object in the GME model database. C++ methods provide convenient read/write access to the objects' properties, attributes, and relations described in GME metamodel.

In the context of GOOSSDM, the lower layers can be conceptualized using levels in GME. The semi-structured data

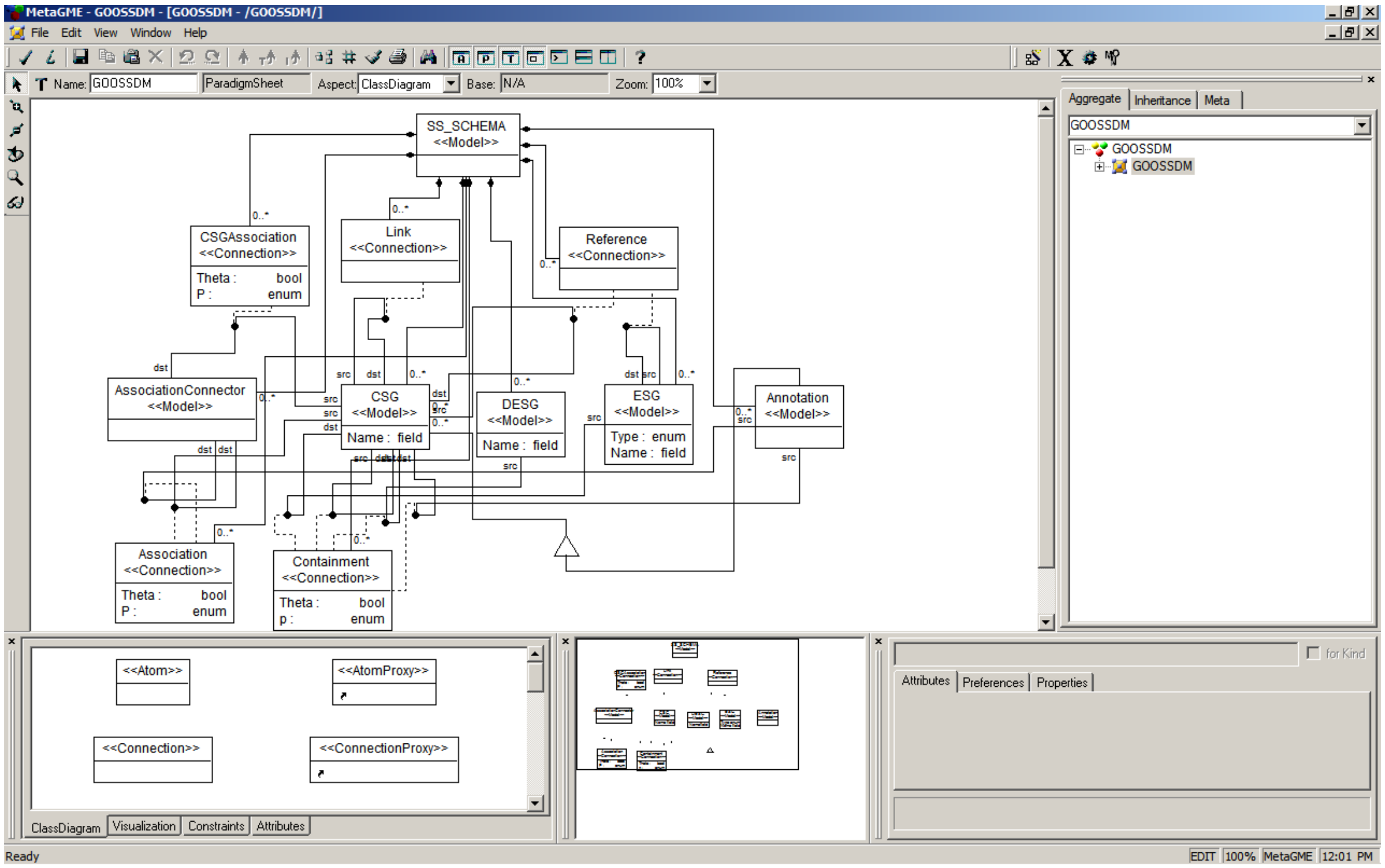

Figure11. Meta-Level Specifications of GOOSSDM model using GME

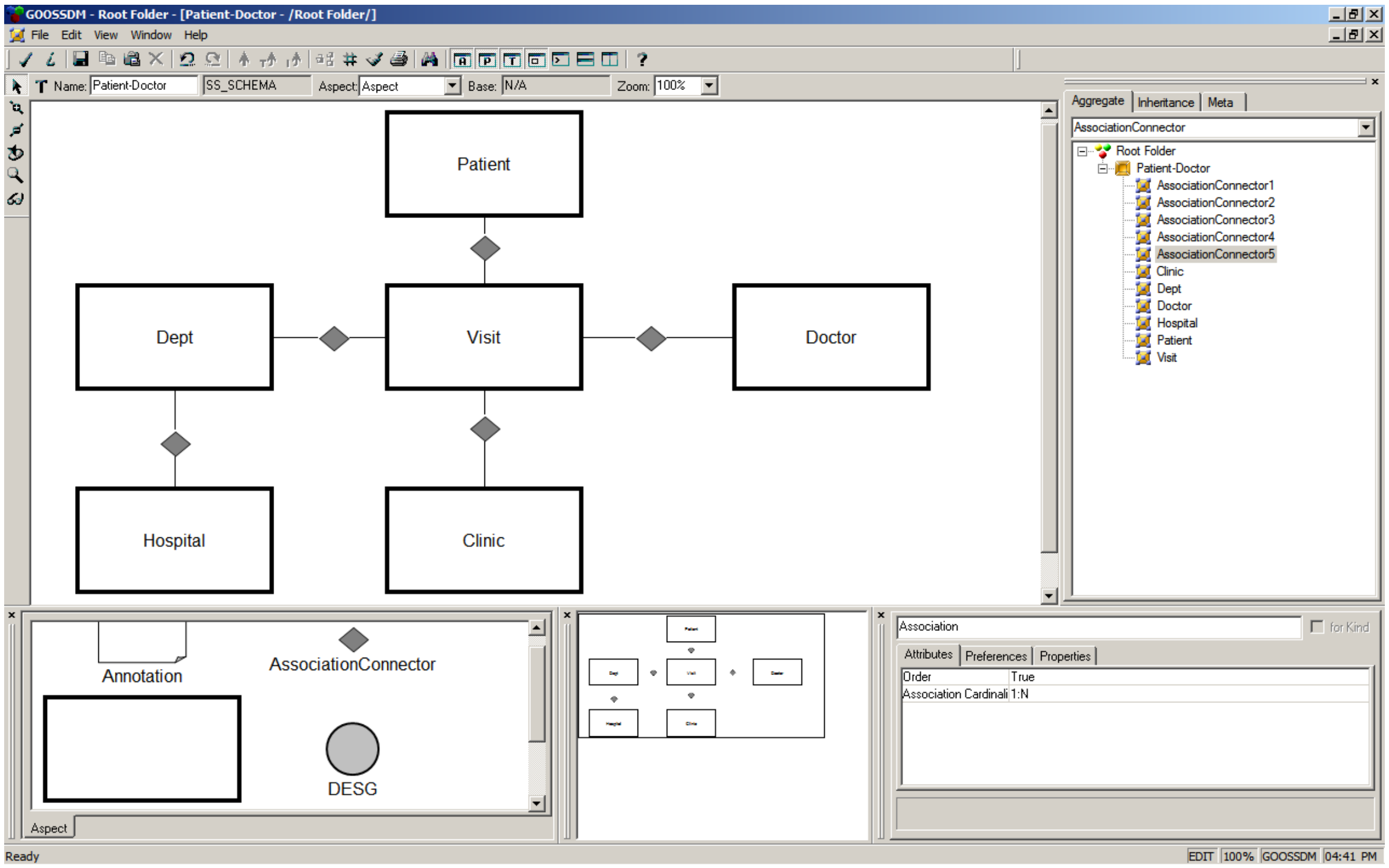

Figure 12. GOOSSDM Schema of *Patient and Doctor* Example using GME

definitions for any given GME model can be configured using meta-level specifications of GOOSSDM. The interpreter will generate the equivalent XML Schema Definitions for any given GME model configured using meta-level specifications of GOOSSDM to represent the semi-structured data at logical level.

The meta-level specifications of GOOSSDM using GME have been shown in Figure 11. The GOOSSDM schema specification of *Patient* and *Doctor* example (Figure 9 and Figure 10) using GOOSSFM meta-level specifications has been shown in Figure 12. The BON based interpreter for GOOSSDM can run from the GME interface to interpret any GOOSSDM schema and to generate the equivalent XSD Code.

## VIII. CONCLUSION

In this paper, a model has been introduced for the conceptual level design of semi-structured data using graph based semantics. This is a comprehensive object oriented conceptual model and the entire semi-structure database can be viewed as a Graph (V, E) in layered organization. The graph based semantics in GOOSSDM model extracts the positive features of both Object and Relational data models and also it maintains the referential integrity inherently. Further the layered organization of the model facilitates to view the semi-structured data schema from different level of abstraction.

The proposed GOOSSDM contains detailed set of semantically enriched constructs and relationships those are necessary to specify the facets of semi-structured database system at conceptual level. Moreover, a set of rules also have been proposed to systematically transform any GOOSSDM schema to its equivalent XSD structure. The expressive powers of the set of transformation rules have been illustrated with suitable examples and case study. Moreover, the proposed model also facilitates the designer to provide alternative design of same schema by changing the ordering scheme, which in result can be transformed in different XSDs with different nesting patterns. It provides better understandability to the users and high flexibility to the designers for creation and / or modification of semi-structured data as well as XML document at conceptual level. The proposed approach is also independent from any implementation issues.

It is also important to note that with the concept of Annotation construct the proposed GOOSSDM facilitate the document – centric design of semi-structured data at conceptual level. Also the proposed model supports irregular, heterogeneous, hierarchical and no-hierarchical structure in data. Moreover, the set of proposed rules are capable to transform systematically the GOOSSDM schema into hierarchical XSD schema. Due to these features, the proposed approach is also capable to design XML document at conceptual level.

The proposed approach also has been automated through the GME based meta-model configuration of GOOSSDM. The meta-level specification of GOOSSDM along with interpreter can be used as a CASE tool for the model by the semi-structured database designer. The tools facilitates the automatic generation of XML Schema Definitions from the conceptual level graphical model, using the set of proposed rule set.

Future studies will concentrate on developing a graphical query language for the proposed approach.

### AUTHORS PROFILE

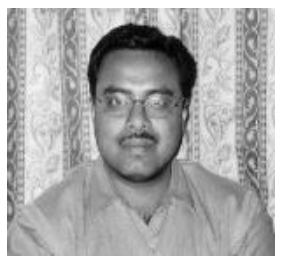

**Anirban Sarkar** is presently a faculty member in the Department of Computer Applications, National Institute of Technology, Durgapur, India. He received his PhD degree from National Institute of Technology, Durgapur, India in 2010. His areas of research interests are Database Systems and Software Engineering. His total numbers of publications in various international platforms are about 25.